\documentclass[12pt]{article}
\usepackage{graphicx}
\usepackage{hyperref}
\usepackage{color}
\usepackage{float}

\usepackage{makeidx}
\usepackage{graphicx}

\usepackage{multicol}
\usepackage[centertags]{amsmath}
\usepackage{amsfonts}
\usepackage{euscript}
\usepackage{subfigure}
\usepackage{color}
\usepackage{amsmath,amssymb,amsthm}

\def\vect#1{\mbox{\boldmath{$#1$}}}

\def\Ac{{\cal A}}

\def\vx{\vec{\vect x}}
\def\vy{\vec{\vect y}}
\def\vz{{ \vec{\vect z}}}

\def\mC{\mathbb{C}}
\def\Cc{{\cal C}}
\newcommand{\cC}{\mathcal{C}}

\def\wG{G}

\newcommand{\bx}{\vect \rho}
\newcommand{\bz}{\vect z}

\def\wg{g}

\newcommand{\eps}{\varepsilon}
\renewcommand{\epsilon}{\eps}
\renewcommand{\leq}{\leqslant}
\renewcommand{\geq}{\geqslant}

\def\bfrho{\mbox{\boldmath$\rho$}}
\def\bfb{\mbox{\boldmath$b$}}

\usepackage{perpage}
\MakePerPage{footnote}
\def\vect#1{\mbox{\boldmath{$#1$}}}
\def\Ac{{\cal A}}
\def\Cc{{\cal C}}
\def\cC{{\cal C}}

\def\bfrho{\mbox{\boldmath$\rho$}}
\def\brho{\mbox{\boldmath$\rho$}}
\def\bfb{\mbox{\boldmath$b$}}
\renewcommand{\bx}{\vect \rho}
\renewcommand{\bz}{\vect z}

\newcommand{\sign}{\text{sign}}
\newcommand{\lc}{\left(}
\newcommand{\rc}{\right)}

\def\mC{\mathbb{R}}

\def\vx{\vec{\vect x}}
\def\vy{\vec{\vect y}}
\def\vz{\vec{\vect z}}
\def\wG{G}
\def\wg{g}
\def\rmi{\mathrm{i}}

\title{The Noise Collector for sparse recovery in high dimensions}

\author{Miguel Moscoso\footnote{Department of Mathematics, Universidad Carlos III de Madrid, Leganes, Madrid 28911, Spain}
, Alexei Novikov\footnote{Department of Mathematics, Pennsylvania State University, University Park, PA 16802}
, George Papanicolaou\footnote{Department of Mathematics, Stanford University, Stanford, CA 94305}
, Chrysoula Tsogka\footnote{Department of Applied Mathematics, University of California, Merced, CA 95343}
}

\begin{document}

\maketitle

\begin{abstract}
The ability to detect sparse signals from noisy high-dimensional data is a top priority in modern science and engineering.
A sparse solution of the linear system $\Ac \vect \rho = \vect b_0$ can be found efficiently with an $\ell_1$-norm minimization approach if the data is noiseless.
Detection of the signal's support from data corrupted by noise is still a challenging problem, especially if the level of noise must be estimated.
We propose a new efficient approach that does not require any parameter estimation. We introduce the Noise Collector (NC) matrix $\cC$ and solve an augmented system
$\Ac \vect \rho + \cC \vect \eta =\vect b_0 +\vect e$, where $\vect e$ is the noise. We show that the $l_1$-norm minimal solution of the augmented system has zero false discovery rate for any level of noise and with probability that tends to one as the dimension of $\vect b_0$ increases to infinity.
We also obtain exact support recovery if the noise is not too large, and develop a Fast Noise Collector Algorithm which makes the computational cost of solving the augmented system comparable to that of the original one. Finally, we demonstrate the effectiveness of the method in applications to passive array imaging.
\end{abstract}


We want to find  sparse  solutions $ \vect \rho \in \mC^K$ for

\begin{equation}
\label{eq:ls}
\Ac \, \vect \rho= \vect b,
\end{equation}
from highly incomplete measurement data $\vect b=\vect b_0 + \vect e \in \mC^N$, corrupted by  noise  $\vect e$ where $ 1 \ll N < K$.  In the noiseless case, $ \vect \rho $
can be found exactly by solving the  optimization problem \cite{Chen01}

\begin{equation}\label{l1normsol1}
\vect \rho_* = \arg\min_{ \small \vect \rho}  \|\vect \rho\|_{\ell_1} , \hbox{ subject to } \Ac \, \vect \rho = \vect b,
\end{equation}
provided the  measurement matrix  $\Ac\in \mC^{N\times K}$ satisfies additional conditions, e.g., decoherence or restricted isometry properties~ \cite{Donoho03,Candes05},
and the solution vector $\brho$  has a small number $M$ of nonzero components or degrees of freedom. 
When measurements are noisy exact recovery is no longer possible. However the exact support of $\vect \rho$ can be determined if the noise is not too strong. The most commonly used approach is  to solve  the $\ell_2$-relaxed form of [\ref{l1normsol1}] 
\begin{align}\label{eq:lasso}
\vect \rho_{\lambda} = 
\arg\min_{ \small \vect \rho} \left(    \lambda  \| \vect \rho \|_{\ell_1} +  \| \vect  \Ac \vect \rho - \vect b  \|^2_{\ell_2}  \right),
\end{align}
known as Lasso in the statistics literature \cite{T}.  There are sufficient conditions for the support of $\vect \rho_{\lambda}$ to be contained within the true support, see e.g. 
Fuchs \cite{Fuchs05}, Tropp \cite{Tropp06} and Wainwright \cite{Wainwright09}. 
These conditions depend 
on the signal-to-noise ratio (SNR), which is not known and must be estimated, and on the regularization parameter $\lambda$, which must be carefully chosen and/or adaptively changed \cite{Zou06}. 
Although such an adaptive procedure improves the outcome, the resulting solutions tend to include a large number of ``false positives" in practice \cite{Sampson13}.
Our contribution is a method for exact support recovery in the presence of additive noise. A key element of this method is that it has no tuning parameters. In particular, 
it does not require any prior knowledge of the level of noise which is often difficult to estimate.

{\bf Main Results.} Suppose $\vect \rho$ is an $M$-sparse solution of the noiseless system in~\eqref{eq:ls},
where the columns of $\Ac$  have unit length. Our main result ensures that we can
recover the support of  $\vect \rho$ by looking at the support of  $\vect \rho_{\tau}$ found as 
\begin{align}\label{rho_t}
\left( \vect \rho_{\tau}, \vect \eta_{\tau} \right) = 
\arg\min_{ \small \vect \rho, \small \vect \eta} \left(    \tau  \| \vect \rho \|_{l_1} +  \| \vect \eta \|_{l_1}  \right),\\
 \hbox{ subject to } \Ac \vect \rho + \Cc \vect \eta =\vect b_0  +  \vect e, \nonumber
\end{align}
with an $O(\sqrt{\ln N})$ weight $\tau$, and an appropriately chosen {\it Noise Collector } matrix  $\cC \in \mC^{N\times \Sigma}$, $\Sigma \gg K$.  
The minimization problem [\ref{rho_t}] can be understood as a relaxation of [\ref{l1normsol1}].
It works by absorbing {\it all} the noise, and possibly some signal, in $\Cc \vect \eta_{\tau}$. 

The following  theorem shows that if the signal is pure noise, and 
the columns of the {\it Noise Collector} are chosen uniformly and independently at random on the unit sphere $\mathbb{S}^{N-1}=\left\{ x \in \mathbb{R}^{N} , \| x \|_{\ell_2} =1  \right\}$, then 
$\Cc \vect \eta_{\tau} =\vect e$ for any level of noise, with high probability. 

Theorem~1 (No phantom signal):  Suppose  $\vect b_0 =0$ and $\vect  e/\| \vect e \|_{l_2}$ is uniformly distributed on  
the unit sphere $\mathbb{S}^{N-1}$. 
Fix $\beta>1$  and draw  $\Sigma=N^{\beta}$  columns for 
$\Cc$  independently from the  uniform distribution on 
$\mathbb{S}^{N-1}$. 
For any $\kappa >0$ there are constants $c_0=c_0(\kappa, \beta)$ and $N_0= N_0(\kappa, \beta)$ 
 such that, for  $\tau = c_0 \sqrt{\ln N}$ and  all $N>N_0$,  $\vect \rho_\tau $, the solution of~\eqref{rho_t}, is zero  with  probability  $1-1/N^{\kappa}$. 

Theorem~1 guarantees a zero false discovery rate in the absence of 
 signals with  meaningful information, with high probability. 
  We generalize this result for the case in which the recorded signals
 carry useful information in the next Theorem, where we show that 
the support of $\vect \rho_\tau$ 
is inside the support of $\vect \rho$.

Theorem~2 (Zero false discoveries): Let $\vect \rho$ be an $M$-sparse solution of the noiseless system $\Ac \vect \rho=\vect b_0$. Assume $\kappa$, $\beta$, the Noise Collector, the noise, and  $\vect \rho_{\tau}$ are the same as in Theorem~1.
In addition, assume that the columns of $\Ac$  are incoherent, in the sense that $|\langle \vect a_i, \vect a_j \rangle| \leq \frac{1}{3M}$.
Then, there are constants $c_0=c_0(\kappa, \beta)$ and $N_0= N_0(\kappa, \beta)$ such that, for  $\tau = c_0 \sqrt{\ln N}$ and all $N>N_0$, 
$\mbox{supp}(\vect \rho_{\tau}) \subseteq \mbox{supp}(\vect \rho)$
with  probability  $1-1/N^{\kappa}$. 

The incoherence conditions in Theorem~2 are needed to guarantee that the true signal does not create false positives elsewhere.  
The next Theorem shows that if the noise is not too large,  then 
$\vect \rho_\tau$ and $\vect \rho$ have exactly the same support.

Theorem~3 (Exact support recovery):  Keep the same assumptions as in Theorem~2. Suppose the magnitudes of the non-zero entries of $\vect \rho$ are bounded by $\gamma$.
 If  $\| \vect e \|_{l_2}/\| \vect b_0 \|_{l_2}\leq c_2/ \sqrt{\ln N}$, $c_2=c_2(\kappa, \beta, \gamma, M)$,   then $\vect \rho_\tau$ and $\vect \rho$ have the same support with  probability
  $1-1/N^{\kappa}$.

{\bf Motivation.} We are interested in imaging accurately sparse scenes using limited and noisy data.
Such imaging problems arise in many areas such as medical imaging \cite{Trzasko09}, structural biology \cite{AlQuraishi11}, radar \cite{Baraniuk07b}, 
and geophysics~\cite{Taylor79}. 
In imaging, the $\ell_1$-norm minimization method in~\eqref{l1normsol1} is used often, in e.g.~\cite{Malioutov05,Romberg08,Herman09,Tropp10,Fannjiang10,CMP13}.
This method has the desirable property of super-resolution, that is, the enhancement  fine scale details of the images using, in this case, prior information about its low dimensional structure (sparsity).
This has been analyzed in different settings by Donoho and Elad \cite{Donoho92}, Cand\`{e}s and Fernandez-Granda \cite{Candes14},
Fannjiang and Liao \cite{Fannjiang12b}, and Borcea and Kocyigit \cite{Borcea15}, among others.  We want to retain this property in our method when the data is 
corrupted by additive noise.

However, noise fundamentally limits the quality of the images formed with  almost all computational imaging techniques. 
Specifically, $\ell_1$-norm minimization produces images that are unstable for low SNR due to the ill-conditioning 
of super-resolution reconstruction schemes. 
 The instability emerges as clutter noise in the images, or {\em grass}, that degrades the resolution. 
 Our initial motivation to introduce the Noise Collector matrix $\cC$ was to regularize the matrix $\Ac$ and, thus, to suppress the clutter in the images.
 We  proposed in~\cite{MNPT-l1}  to seek the minimal $\ell_1$-norm solution of the augmented linear system $\Ac \bfrho + \Cc \vect \eta=\bfb$.
The idea was to choose the columns of $\cC$ almost  orthogonal to those of $\Ac$. 
Indeed, the condition number of  $[\Ac \, | \,  \Cc]$ 
becomes $O(1)$ when $O(N)$ columns of  $\Cc$ are taken at random. This essentially follows from the bounds on the largest and the smallest nonzero singular values of random matrices, 
 see e.g. Theorem 4.6.1 in~\cite{vershynin}.  

The idea to create a dictionary for noise is not new. For example,  the work by Laska {\em et al.} \cite{Laska09} considers a 
 specific version of the measurement noise model so $ \vect b = \Ac \vect \rho +\cC \vect e$, where $\cC$ is a matrix with fewer (orthonormal) 
 columns than rows, and the noise vector $\vect e$ is sparse. $\cC$ represents the basis in which the noise is sparse
and it is assumed to be known.  Then, they show that it is possible to recover sparse signals and sparse noise exactly using 
$\ell_1$-norm minimization algorithms. We stress that we do not assume here that the noise is sparse. In our work, 
 the noise is large (SNR can be small) and is evenly distributed across the data, so it cannot be sparsely accommodated.

 To  suppress the clutter, our theory in~\cite{MNPT-l1} required exponentially many columns, so $\Sigma \lesssim e^N$. 
 This seemed to make the noise collector impractical, but the numerical experiments suggested that $O(N)$ columns were enough to obtain excellent results.
We address this issue here and explain why the noise collector matrix $\cC$ only needs algebraically many columns. 
Moreover, to make the absorption of noise less expensive, and thus improve the algorithm in~\cite{MNPT-l1}, we introduce the weight $\tau$ in~\eqref{rho_t}. 
Indeed, by weighting the columns of the noise collector matrix $\cC$ with respect to those in the model matrix $\Ac$,  the algorithm
 now produces images with no clutter at all, no matter how much noise is added to the data.
 

Finally, we want the {\em Noise Collector} to be efficient, with almost no extra computational cost with respect to the Lasso problem in [\ref{eq:lasso}].
To this end, it is constructed using circulant matrices that allows for efficient matrix vector multiplications using FFTs. 

The proofs of Theorems~1,~2, and~3 are given in Section Proofs.  We now explain how the {\em Noise Collector}  works.

\section*{The Noise Collector}\label{sec:theory}
\vspace{-0.15cm}
The construction of the {\em Noise Collector} matrix $\Cc$ starts with the following three key properties. Firstly, its columns should be sufficiently orthogonal to the columns of $\Ac$,
 so  it does not absorb signals with "meaningful" information. Secondly, the
columns  of $\Cc$  should be uniformly distributed on the unit sphere $\mathbb{S}^{N-1}$  so that we could approximate well a typical noise vector. Thirdly, 
the  number of columns in $\Cc$ should grow slower than exponential with $N$, otherwise the method is impractical. 
One way to guarantee all three properties is to impose
  \begin{equation}\label{deco_all}
   |\langle \vect a_i,\vect c_j \rangle | < \frac{\alpha }{\sqrt{N}}~\forall i,j\,, \hbox{ and }
   |\langle \vect c_i,\vect c_j \rangle | <\frac{\alpha }{\sqrt{N}}~ \forall  i \neq j, 
\end{equation}
with $\alpha>1$, and fill out $\Cc$ drawing $\vect c_i$ at random with rejections until the rejection rate becomes too high. Then, by construction, the columns of $\Cc$ are almost orthogonal
 to the columns of $\Ac$, and when the rejection rate becomes too high this implies that we can not pack more N-dimensional unit vectors into $\Cc$ and, thus, we  
can approximate well a typical noise vector.
Finally, the Kabatjanskii-Levenstein inequality  (see discussion in~\cite{tao})
implies that 
the number $\Sigma$ of columns in $\Cc$ grows at most polynomially:
$ \Sigma  \leq N^{\alpha^2}$.

It is, however, more convenient for the proofs to use a probabilistic version of~[\ref{deco_all}].  Suppose that  the columns of $\Cc$ are drawn at random independently. 
Then, the dot product of any
 two random unit vectors is still typically of order $1/\sqrt{N}$, see e.g.~\cite{vershynin}. If the number of columns grows polynomially,  we only 
have to sacrifice an asymptotically 
negligible event where our {\em Noise Collector} does not 
satisfy the three key  properties, and the decoherence constraints in~[\ref{deco_all}] are weakened by a logarithmic factor.  The next Lemma is proved in Section 
Proofs.

Lemma~1:  Suppose $\Sigma=N^{\beta}$, $\beta > 1$, vectors $ \vect c_i \in \mathbb{S}^{N-1}$ are drawn at random and independently. Then,  (i) for any $\kappa>0$ there are constants $c_0(\kappa,\beta)$ and $\alpha>1/2$, such that
\vspace{-.15cm}
\begin{equation}\label{deco_2}
\forall i,j  |\langle \vect a_i,\vect c_j \rangle | < c_0 \sqrt{\ln N} /\sqrt{N},
\end{equation}
and (ii) for any $\vect e \in \mathbb{S}^{N-1}$  there exists at least one $\vect c_j $, so that
\vspace{-.15cm}
\begin{equation}\label{deco_1}
  |\langle \vect e,\vect c_j \rangle | \geq  \alpha/\sqrt{N}\, 
\end{equation}
with probability $1-1/N^{\kappa}$. 

The estimate in~[\ref{deco_2}] implies that  any solution 
$\cC \vect \eta = \vect a_i$
satisfies, for any $i \leq N$,
\begin{equation}\label{l1_bound}
\| \vect \eta \|_{\ell_1} \geq \frac{\sqrt{N}}{c_0 \sqrt{\ln N}}\, ,
\end{equation}
 with probability $1-1/N^{\kappa}$. This estimate measures how expensive it is to approximate columns of $\Ac$ with the {\em Noise Collector}. 
 In turn, the weight $\tau$ should be chosen so that it is expensive to approximate noise using columns of $\Ac$. It cannot be taken too large, though, because we may loose the signal. 
 In fact, one can prove that if $\tau \geq \sqrt{N}/\alpha$, then $\vect \rho_{\tau} \equiv 0$ for any $\vect \rho$ and any level of noise. 
 Intuitively, the weight $\tau$ characterizes the rate at which the signal is lost as the noise increases. 
 \begin{figure}[htbp]
\begin{center}
\includegraphics[scale=0.26]{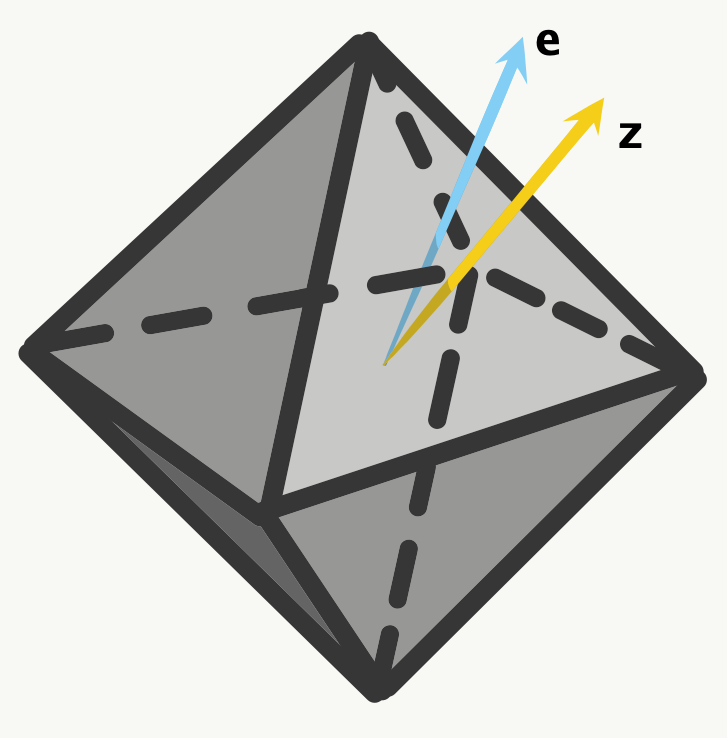} 
\includegraphics[scale=0.27]{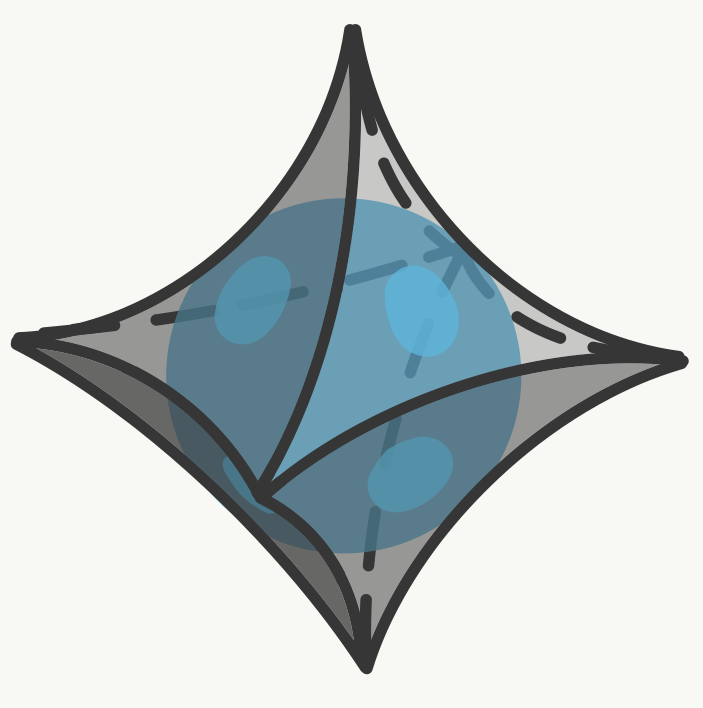}
\caption{Left: Illustration of the $\ell_1$ unit cube and the map $\Phi_{\Cc}$. Right: V.Milman's hyperbolic view of the $l_1$ ball of radius $O(\sqrt{N})$ and 
$\mathbb{S}^{N-1}$.}
\label{phi_map}
\end{center}
\end{figure}

To explain the theoretical lower bound $\tau \geq c_0 \sqrt{\ln N}$ we turn to the geometric interpretation of duality in linear programming. 
Suppose $\tau=\infty$ and there is no signal, $\vect b_0$.
Then,  the solution of~[\ref{rho_t}] satisfies $(\vect \rho_{\infty},  \vect \eta_{\infty})=(\vect 0, \vect \eta)$,
 and there is  a {\it dual certificate}  $\vect z$  of optimality 
of $(\vect 0, \vect \eta)$ for $\tau=\infty$
that satisfies
\[
\langle \vect c_j, \vect  z \rangle = \hbox{sign}(\eta_j)\, \hbox{ if } \eta_j \neq 0, \hbox{ and } | \langle \vect c_j, \vect  z \rangle|  \leq 1\, \hbox{ if } \eta_j =0. 
\]
Define  a nonlinear map 
\begin{equation}\label{map_Phi}
\Phi_\Cc:  \vect e \to \vect z,
\end{equation} 
where $\vect e \in  \mathbb{R}^{N}$ is the  noise vector in~[\ref{rho_t}], and $\vect z$ is the dual certificate of  optimality of $(\vect 0, \vect \eta)$ for $\tau=\infty$. 
For example, if $\Cc$ is the identity matrix, then $\Phi_\Cc( \vect e)= (\hbox{sign}(e_{1}), \dots, \hbox{sign}(e_{N}))$; see Figure~\ref{phi_map}-left.
 If  $\vect z=\Phi_\Cc( \vect e)$  remains  a  dual certificate of  optimality of $(\vect 0, \vect \eta)$ for  $\tau =c_0 \sqrt{\ln N}$, then it implies
that support$(\vect \rho_\tau) \subset $  support$(\vect \rho)$ for such $\tau$. Thus, Theorem~1 follows once we check that
 \begin{equation}\label{no_signal_2}
  |\langle \vect a_j, \vect z \rangle| < \tau, \hbox{ for all } j \leq K,
  \end{equation}
holds with large probability. Thus, we need to understand the statistics of $\vect z= \Phi_\Cc( \vect e)$, given that $\vect e/\| \vect e \|$ is uniformly distributed on 
$\mathbb{S}^{N-1}$. The columns of the {\em Noise Collector} were also uniformly distributed on $\mathbb{S}^{N-1}$, thus the vector $\vect   n= \vect z/\| \vect z\|_{l_2}$ 
has to be 
uniformly  distributed on $\mathbb{S}^{N-1}$ as well.  The chance~\eqref{no_signal_2}  does not hold, could be estimated by the area of the intersection of 
 the unit sphere   $\mathbb{S}^{N-1}$ and the $l_1$ ball of radius $O(\sqrt{N})$ (see Figure~\ref{phi_map}-right), which can be shown to be small 
by standard estimates from high-dimensional probability.

By construction, the columns of the combined matrix $[\Ac \, | \,  \Cc]$ are incoherent. This is the key observation, that allows us to prove Theorems~2 and~3 using
standard techniques, see e.g.~\cite{MNPT-l1}. 
In particular, we automatically have  exact recovery  by the standard arguments~\cite{Donoho03} applied to   $[\Ac  \, | \, \Cc]$ if the data is noiseless.

Lemma 2 (Exact Recovery): Suppose $\vect \rho$ is an $M$-sparse solution of $\Ac \vect \rho=\vect b$, and    there is no noise, $\vect e =0$. 
In addition, assume that the columns of $\Ac$  are incoherent: $|\langle \vect a_i, \vect a_j \rangle| \leq \frac{1}{3M}$.
Then, the solution to~[\ref{rho_t}] satisfies $\bfrho_\tau=\bfrho$ for all 

\begin{equation}\label{req}
M < \frac{\sqrt{N}}{ c_0 \sqrt{\ln N} \tau}\, ,
\end{equation}
with  probability $1-1/N^{\kappa}$.


\section*{Fast Noise Collector Algorithm}
\label{sec:algorithm}
To find the minimizer [\ref{rho_t}], we consider a variational approach. We define the function
\begin{eqnarray} 
&& F(\bx, \vect \eta, \bz) = \lambda\,(\tau \| \bx \|_{\ell_1} +  \| \vect \eta \|_{\ell_1}) \\
&& + \frac{1}{2} \| \Ac  \bx  + \Cc  \vect \eta - \bfb \|^2_{\ell_2} + \langle \bz, \bfb - \Ac \bx - \Cc  \vect \eta \rangle \nonumber
\end{eqnarray}
for a weight $\tau=c_0\sqrt{\ln N}$, and determine the solution as
\begin{equation}\label{min-max}
\max_{\bz} \min_{\bx,\vect \eta} F(\bx,\vect \eta,\bz) .
\end{equation}
The key observation is that this variational principle finds the minimum in [\ref{rho_t}] exactly for all values of the regularization parameter $\lambda$.
Hence, the proposed method is fully automated, meaning that it has no tuning parameters. To determine the exact extremum in [\ref{min-max}], we use the iterative soft thresholding algorithm GeLMA~\cite{Moscoso12}
 that works as follows .
 
 For  $\beta=1.5$  we use $\tau= 0.8 \sqrt{\ln N}$ in our numerical experiments. For optimal results, 
 one can calibrate $c_0$ to be the smallest constant such that Theorem 1 holds, that is, we see no phantom signals when the algorithm is fed with pure noise.

Pick a value for the regularization parameter $\lambda$, e.g. $\lambda=1$. Choose step sizes $\Delta t_1< 2/\|[\Ac \, | \, \Cc]\|^2$ and 
$\Delta t_2< \lambda/\|\Ac\|$ \footnote{Choosing two step sizes instead of the smaller one $\Delta t_1$ improves the convergence speed.}. Set $\vect \rho_0= \vect 0$, $\vect \eta_0=\vect 0$, $\vect z_0=\vect 0$, and
iterate for $k\geq 0$:
\begin{eqnarray} 
&& \vect r = \vect b - \Ac \,\bfrho_k - \Cc \,\vect\eta_k\nonumber \, ,\\
&&\vect \rho_{k+1}=\mathcal{S}_{ \, \tau \, \lambda \Delta t_1} \lc \bfrho_k +\Delta t_1 \, \Ac^*(\vect z_k+ \vect r)\rc
\nonumber \, ,\\
&&\vect \eta_{k+1}=\mathcal{S}_{\lambda \Delta t_1} \lc \vect\eta_k +\Delta t_1  \, \Cc^*(\vect z_k+ \vect r)\rc
\nonumber \, ,\\
&&\vect z_{k+1} = \vect z_k + \Delta t_2 \, \vect r \label{eq:algo}\, ,
\end{eqnarray}
where  $\mathcal{S}_{ \lambda}(y_i)=\sign(y_i)\max\{0,|y_i|-\lambda\}$. 

The {\em Noise Collector} matrix $\cC$ is computed by drawing $N^{\beta-1}$ normally distributed $N$-dimensional vectors, normalized to unit length. 
These are the generating vectors of the {\em Noise Collector}. From each of them a circulant $N\times N$ matrix $\cC_i$, $i=1,\ldots,N^{\beta-1}$, is constructed. The {\em Noise Collector} matrix is obtained by concatenation, 
so $\Cc = \left[ \Cc_1 \left| \Cc_2 \left|  \ldots \left| \Cc_{N^{\beta-1}} \right. \right. \right. \right]$. Exploiting the circulant structure of the matrices $\Cc_i$, we perform the matrix vector multiplications $\Cc \vect  \eta_k$ and $\Cc^* (\vect z_k + \vect r)$ in \eqref{eq:algo} using the FFT \cite{Gray06}. This makes the complexity associated to the {\em Noise Collector}  $O(N^{\beta} \log(N))$. Note that only the $N^{\beta-1}$ generating vectors are stored, and not the entire $N \times N^\beta$ {\em Noise Collector} matrix. In practice, we use $\beta \approx 1.5$ which makes the cost of using the {\em Noise Collector} negligible, as typically $K \gg N^{\beta-1}$. 

\section*{Application to imaging}
\label{sec:imaging}
\begin{figure}[htbp]
    \centering
    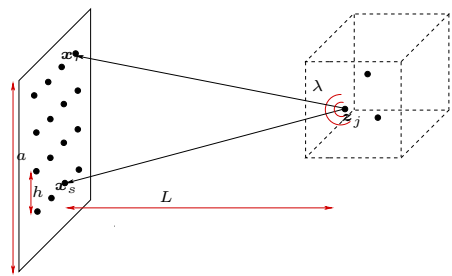
        \caption{General setup for passive array imaging. The source at $\vz_j$  emits a signal that is recorded at all array elements $\vx_r$, $r=1,\ldots,N$.}
    \label{fig:setup}
    \end{figure}

We consider passive array imaging of point sources. The problem consists in determining the positions $\vz_{j}$ and the complex\footnote{We chose to work with real numbers in the previous sections for ease of presentation but the results also hold with complex numbers.} amplitudes $\alpha_j$,
$j=1,\dots,M$, of a few point sources from measurements of polychromatic signals on an array of receivers; see Figure \ref{fig:setup}. 
The imaging system is characterized by the array aperture $a$, the distance $L$ to the sources, the bandwidth $B$ 
and the central wavelength $\lambda_0$.      

The sources are located inside an image window IW, which is discretized with a uniform grid of points $\vy_k$, $k=1,\ldots,K$. 
The unknown is the source vector 
$\vect \rho=[\rho_{1},\ldots,\rho_{K}]^{\intercal}\in\mathbb{C}^K$,
whose components $\rho_k$ correspond to the complex amplitudes of the $M$ sources at the grid points $\vy_k$, $k=1,\ldots,K$, with $K\gg M$. For the true source vector we have
$\rho_{k} = \alpha_j$ if  $\vy_{k} = \vz_j$ for some  $j=1,\ldots,M$, while $\rho_{k} =0$ otherwise. 

Denoting by $\wG(\vx,\vy;\omega)$ the Green's function for the propagation of a 
signal of angular frequency $\omega$ from point $\vy$ to point $\vx$, we define the single-frequency 
Green's function vector that connects a point $\vy$ in the IW with all points $\vx_r$, $r = 1,\ldots,N$, on the array as
$$
\vect \wg(\vy;\omega)=[\wG(\vx_{1},\vy;\omega), \wG(\vx_{2},\vy;\omega),\ldots,
\wG(\vx_{N},\vy;\omega)]^{\intercal} \in \mathbb{C}^{N} \,.
$$
In a homogeneous medium in three dimensions,
$
\wG(\vx,\vy;\omega)=  \frac{\exp\{\rmi \omega|\vx-\vy|/c_0\}}{4\pi|\vx-\vy|}   
$.

The data for the imaging problem are the signals 
\begin{equation}
\label{response}
b(\vx_r, \omega_l)=\sum_{j=1}^M\alpha_j  \wG (\vx_r,\vz_j; \omega_l)
\end{equation}
recorded at receiver locations $\vx_r$, $r = 1,\ldots,N$, at frequencies $\omega_l$, $l=1,\dots,S$.
These data are stacked in a column vector 
\begin{equation}
\vect b = [ \vect  b(\omega_1)^{\intercal},\vect  b(\omega_2)^{\intercal},\dots, \vect b(\omega_S)^{\intercal}]^{\intercal}
\in \mathbb{C}^{(N\cdot S)}\,,
\end{equation}
with
$
\vect b(\omega_l) =  [ b(\vx_1, \omega_l), b(\vx_2, \omega_l),\dots,b(\vx_N, \omega_l)]^{\intercal}\in \mathbb{C}^{N}
$. Then, 
$\Ac\,\bfrho = \bfb$,
with $ \Ac$ the $(N\cdot S)\times K$ measurement matrix whose columns $\vect a_k$ are the multiple-frequency Green's function vectors
\begin{equation}\label{eq:ak}
\vect a_k   = 
[ \vect\wg(\vy_k ; \omega_1)^{\intercal},  \vect\wg(\vy_k; \omega_2)^{\intercal}, \dots, \vect\wg(\vy_k; \omega_{S})^{\intercal} ]^{\intercal} \in \mathbb{C}^{(N\cdot S)}\, ,
 \end{equation}
normalized to have length one.
The system $\Ac\,\bfrho = \bfb$ relates the unknown vector  $\bfrho\in \mathbb{C}^K$ to the data vector $\bfb \in\mathbb{C}^{(N\cdot S)}$.

Next, we illustrate the performance of the Noise Collector in this imaging setup. The most important features  are that (i) no calibration is necessary with respect to the level of noise, (ii)  exact support recovery for relatively large levels of noise (i.e., $\| \vect e \|_{l_2}/\| \vect b_0 \|_{l_2}\leq c_2/ \sqrt{\ln N}$), and (iii)
zero false discovery rate for all levels of noise, with high probability.

We consider a high frequency microwave imaging regime with central frequency $f_0=60$GHz corresponding to $\lambda_0=5$mm. We make measurements for $S=25$ equally spaced frequencies spanning a bandwidth $B=20$GHz. The array has $N=25$ receivers and an aperture $a=50$cm. The distance from the array to the center of the imaging window is $L=50$cm. Then, the resolution 
is $\lambda_0 L/a=5$mm in the cross-range (direction parallel to the array) and $c_0/B=15$mm in range (direction of propagation). These parameters are typical in microwave scanning technology \cite{Laviada15}. 

\begin{figure}[htbp]
\begin{tabular}{cc}
\includegraphics[scale=0.34]{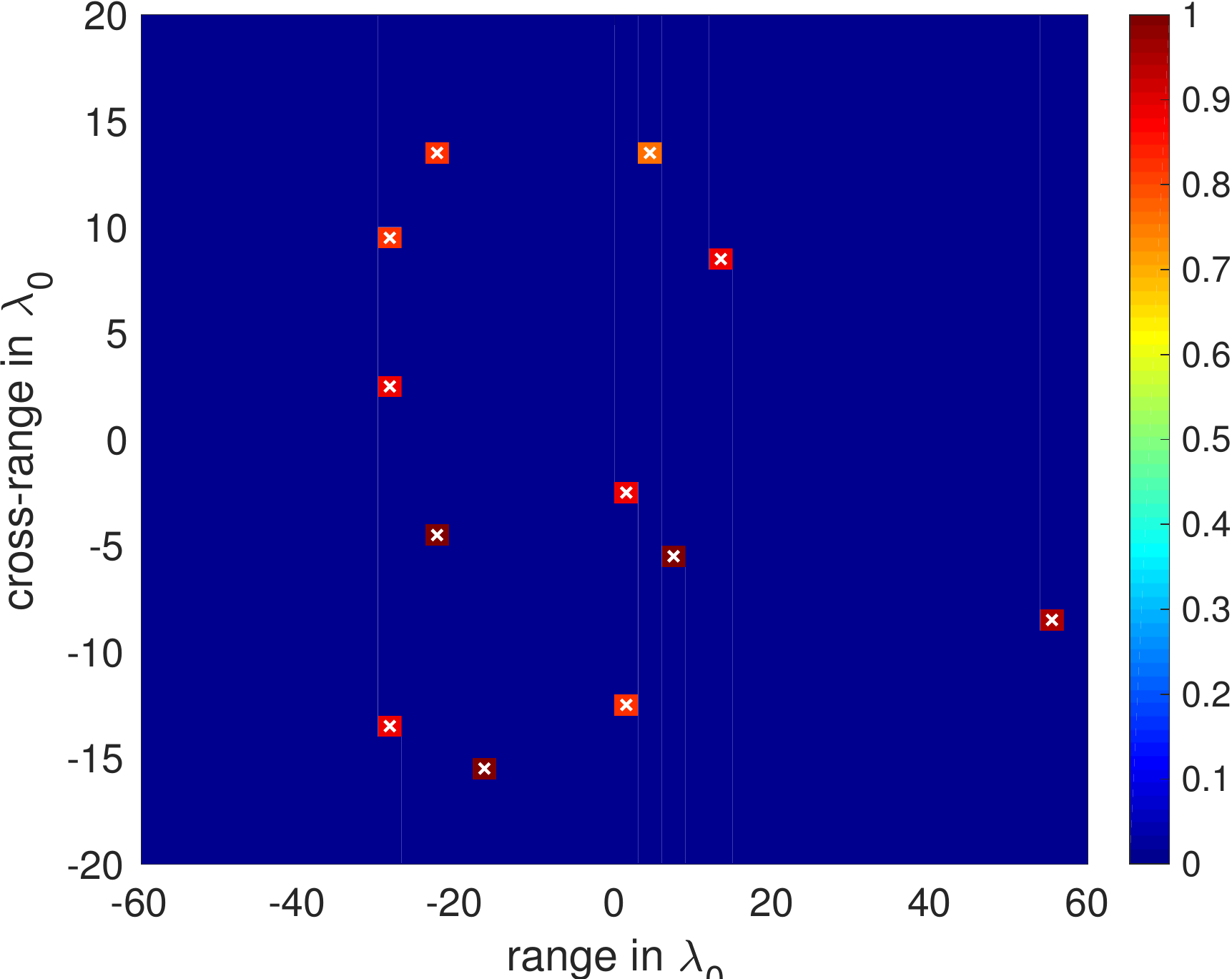} & 
\raisebox{0.15cm}{
\includegraphics[scale=0.365]{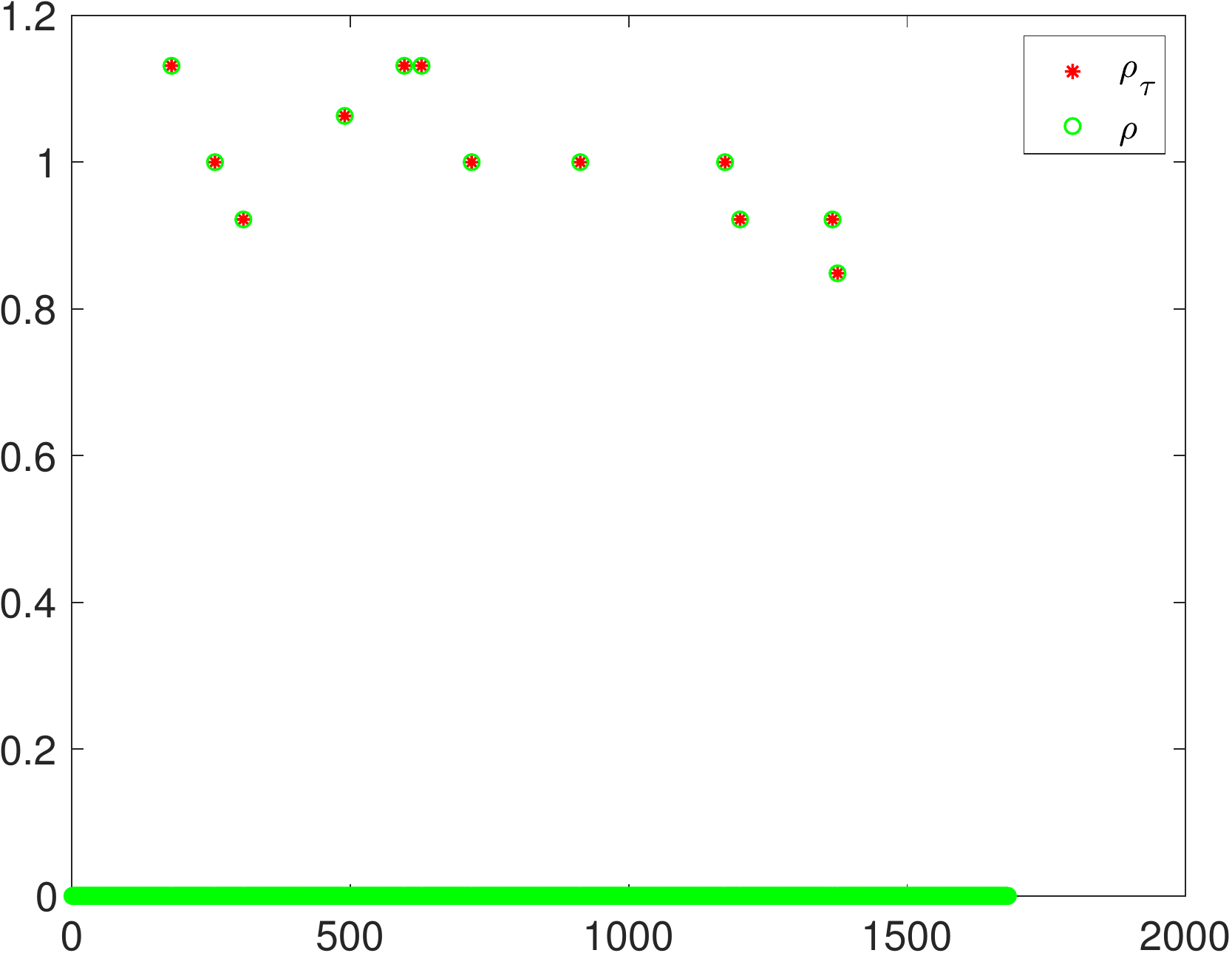} }\\
\end{tabular}
\caption{Noiseless data. The exact solution is recovered for any value of $\tau$ in [\ref{eq:algo}];  Left: the true image. Right: the recovered solution vector, $\rho_\tau$, is plotted with red stars and the true solution vector, $\rho$,  
with green circles.}
\label{fig0}
\end{figure}

 We seek to image a source vector with sparsity $M=12$; see the left plot in Fig. \ref{fig0}. The size of the imaging window is 20cm$\times$60cm and the pixel spacing is 5mm$\times$15mm.  The number of unknowns is, therefore, $K=1681$ and the number of data is $NS=625$. The size of the noise collector is taken to be $\Sigma=10^4$, so $\beta\approx 1.5$.  When the data is noiseless, we obtain exact recovery as expected; see the right plot in Fig. \ref{fig0}

\begin{figure*}[htbp]
\begin{center}
\begin{tabular}{cccc}
 no NC& with NC and $\tau=1$&with NC and $\tau=2$ & $\ell_2$ on the support\\
\includegraphics[scale=0.21]{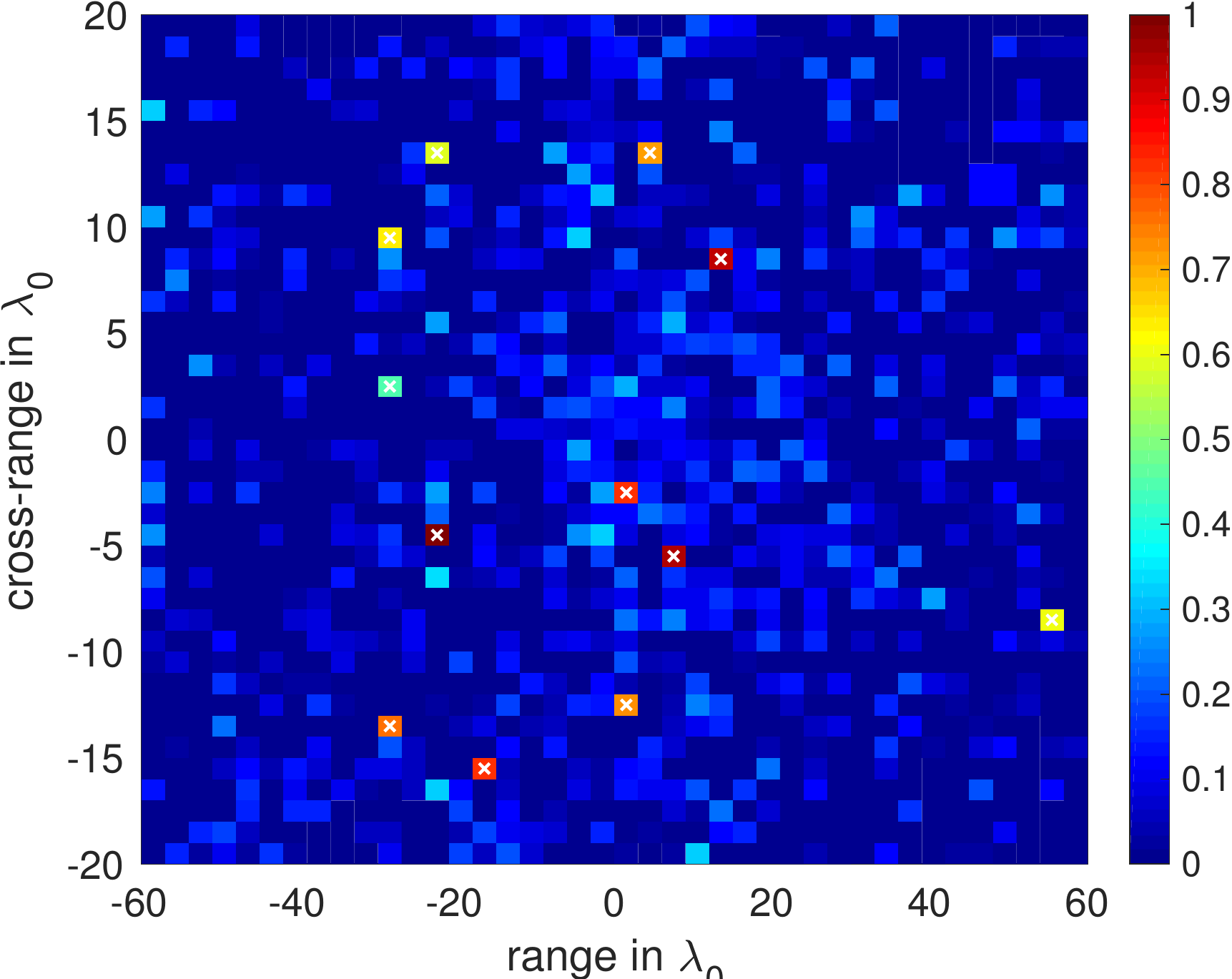}&
\includegraphics[scale=0.21]{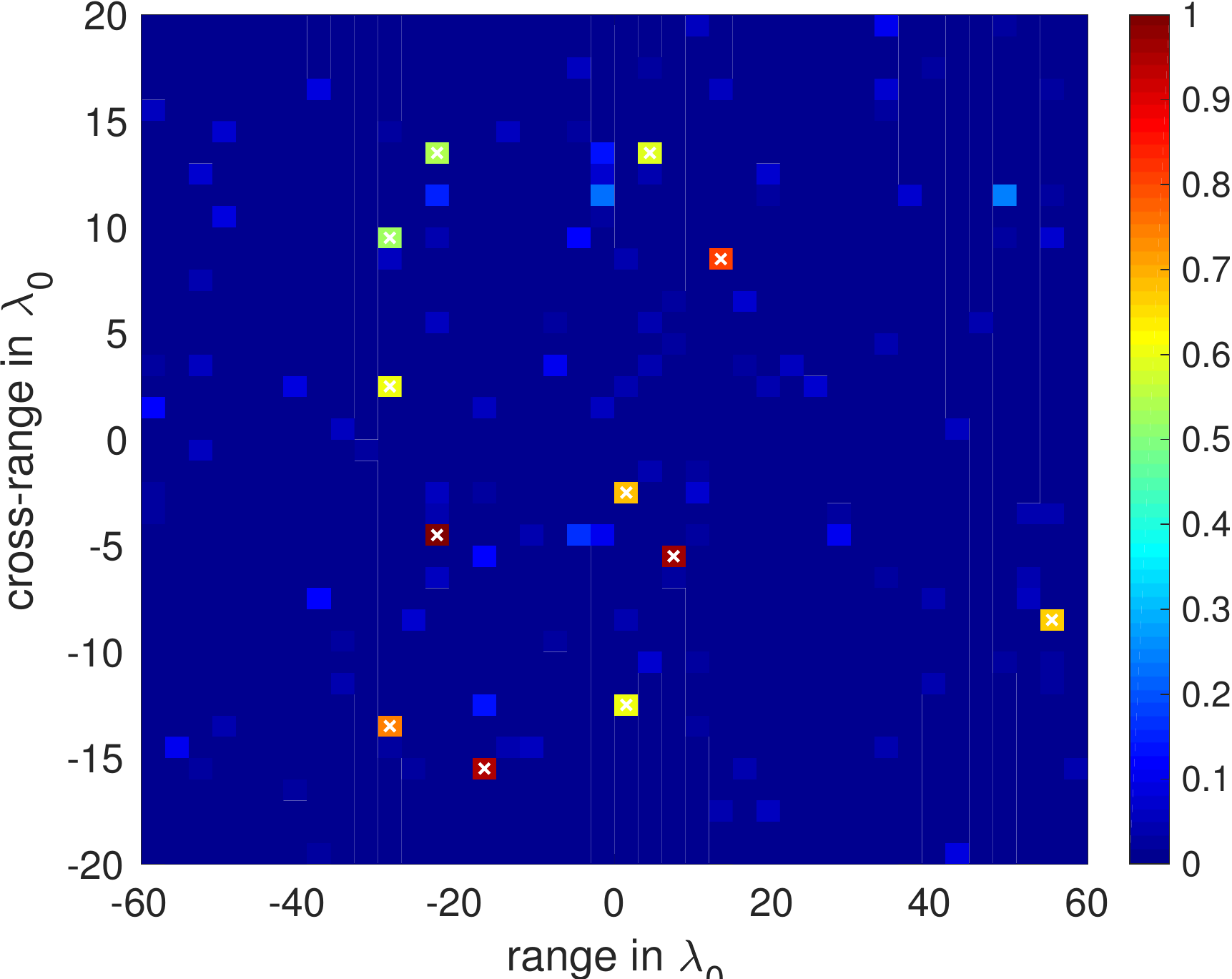} &
\includegraphics[scale=0.21]{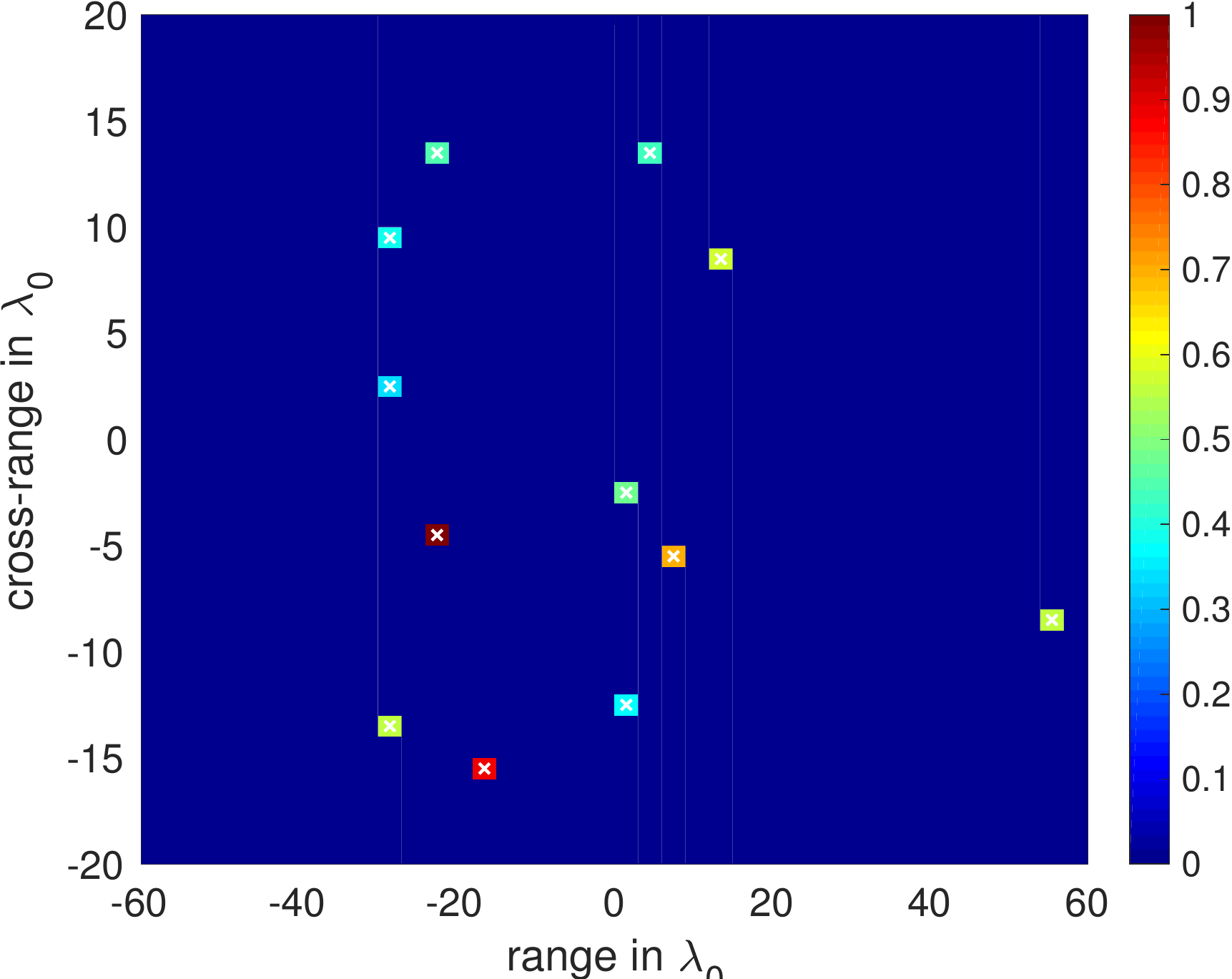} &
\includegraphics[scale=0.21]{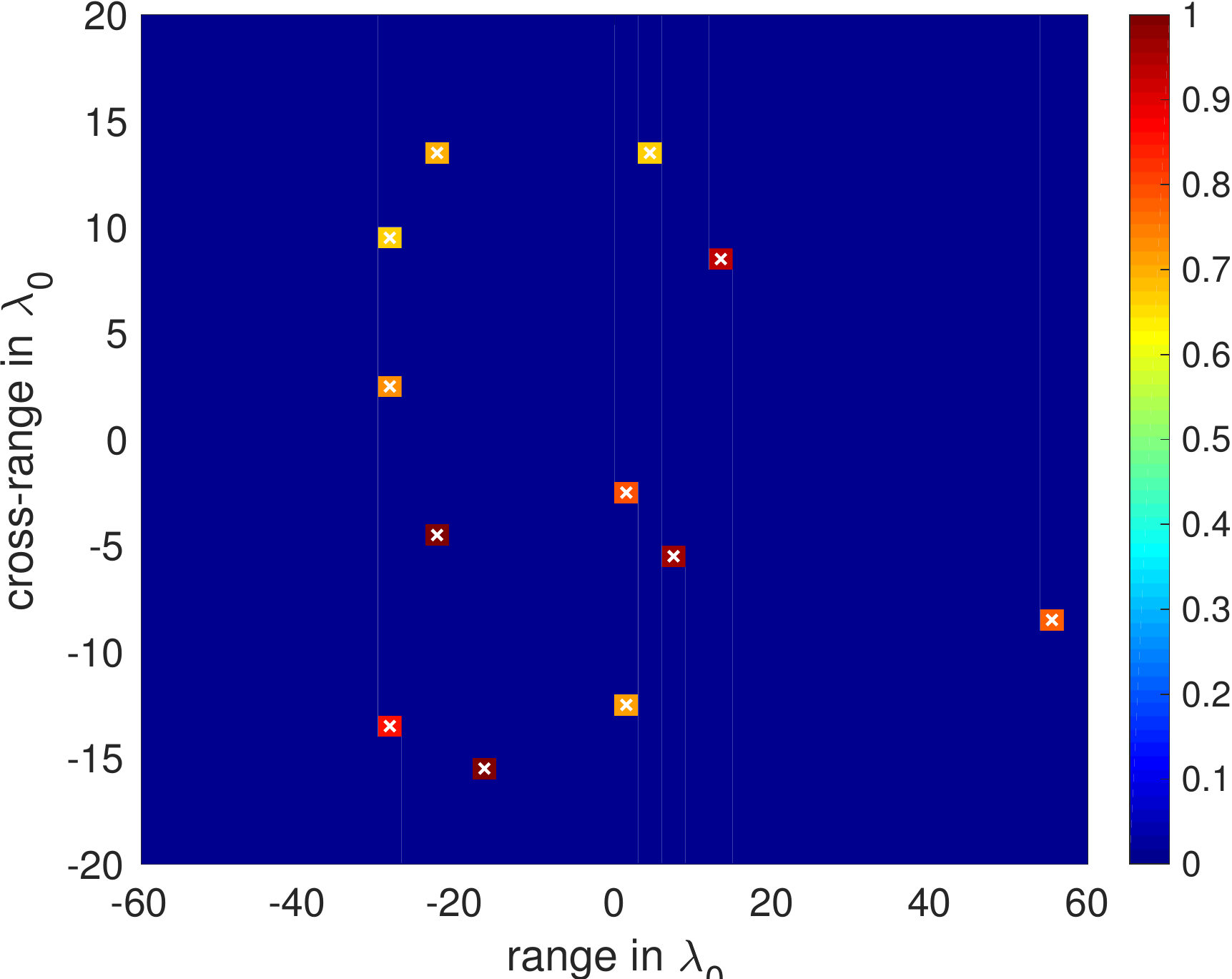} \\
\includegraphics[scale=0.21]{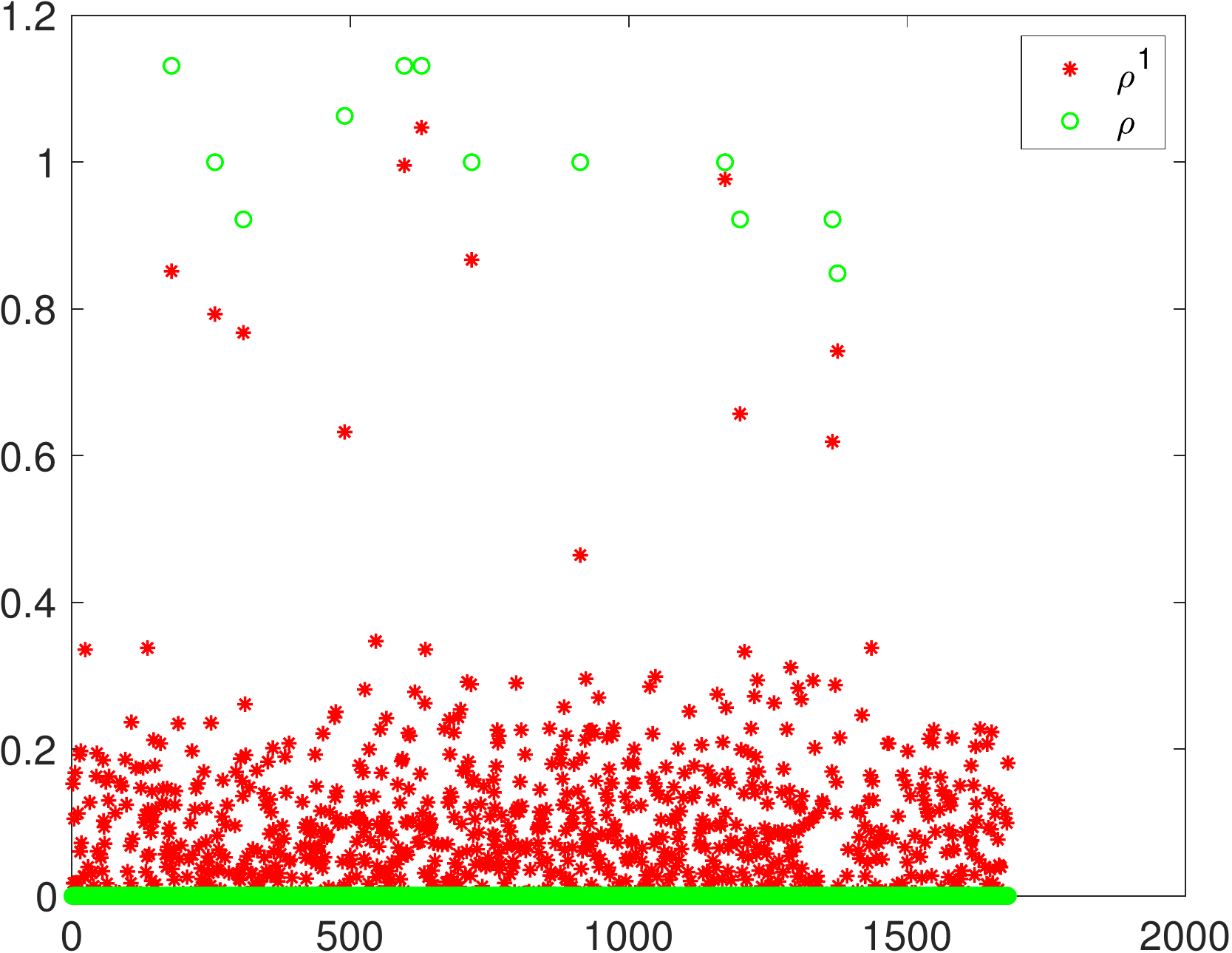}&
\includegraphics[scale=0.21]{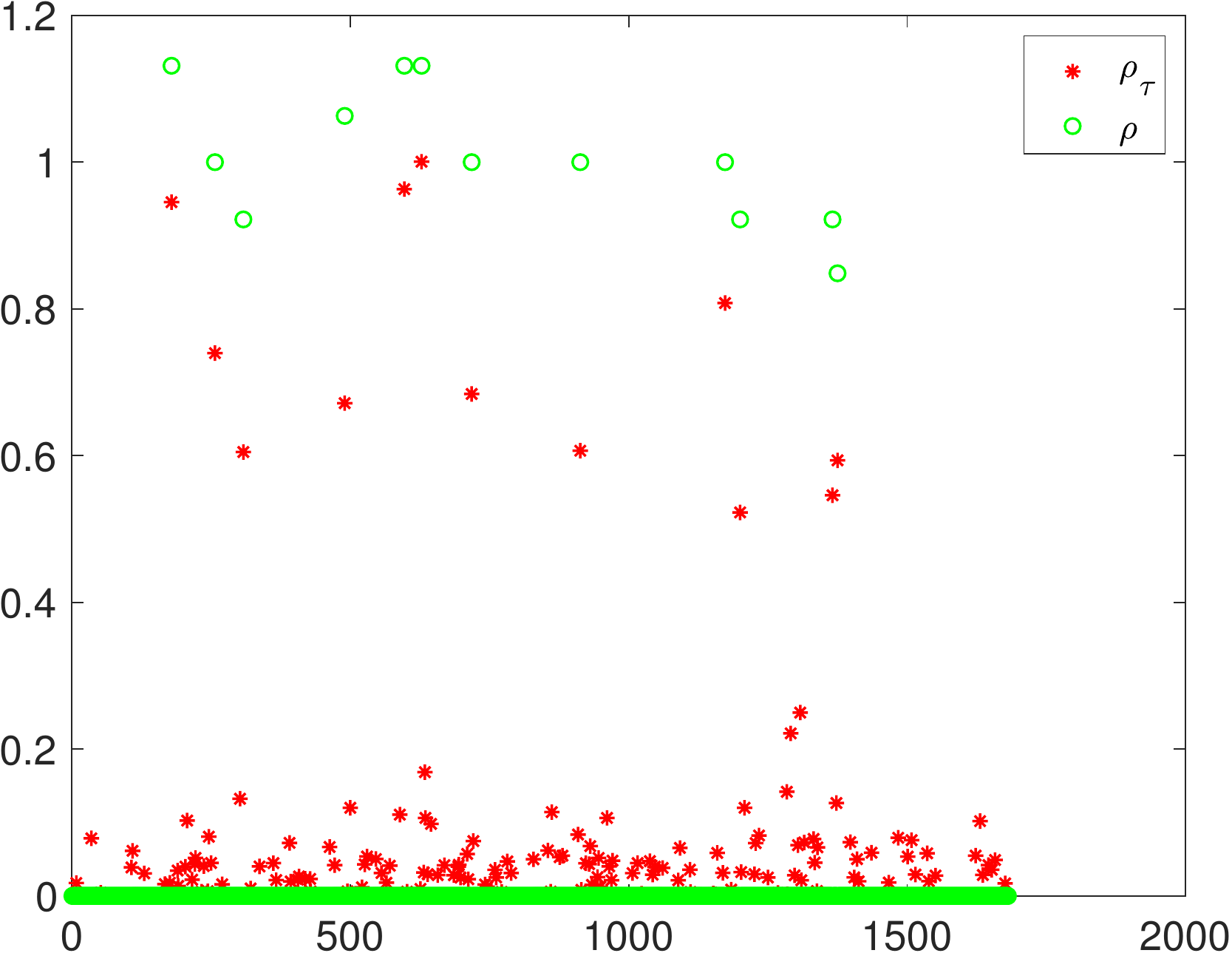} &
\includegraphics[scale=0.21]{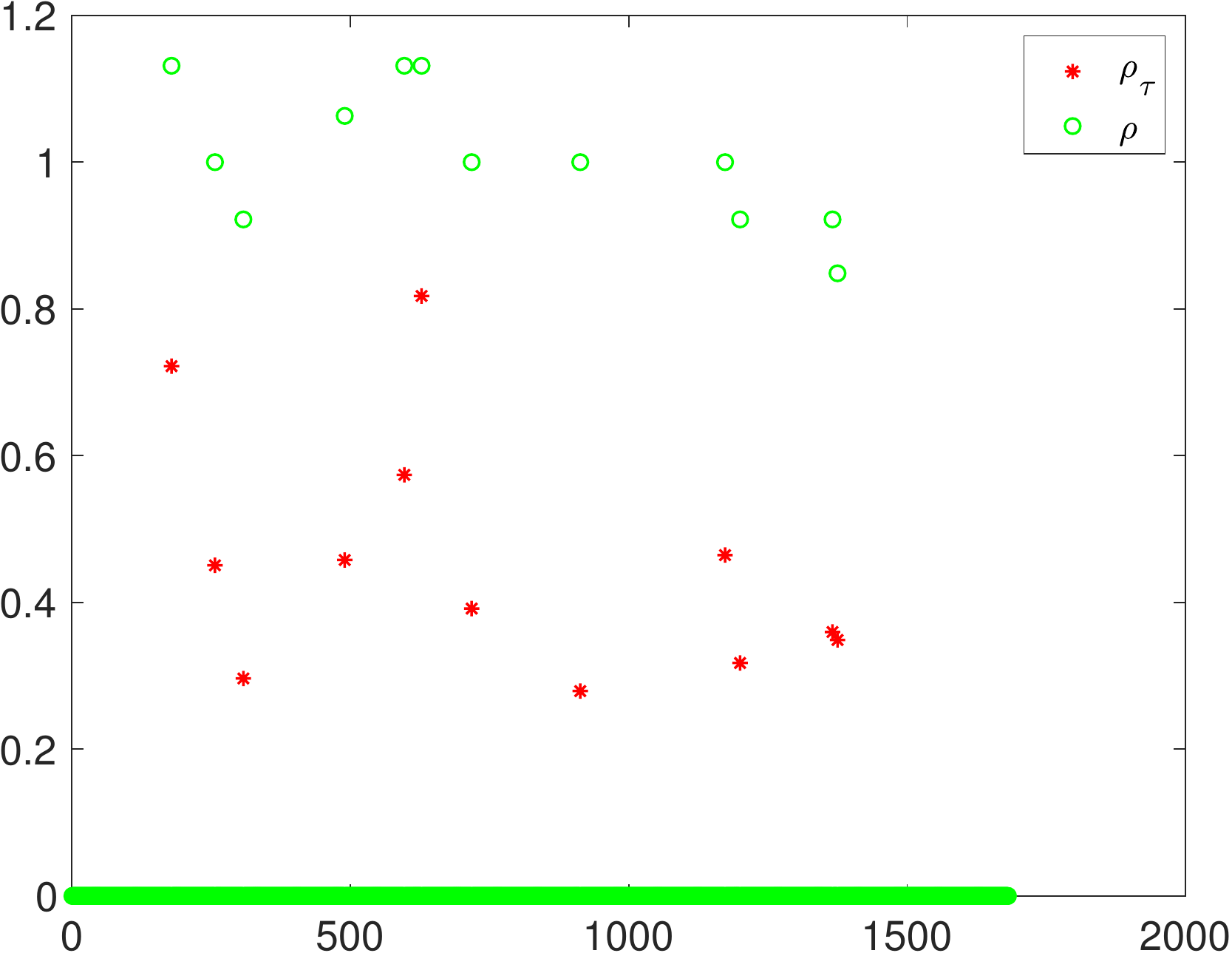}&
\includegraphics[scale=0.21]{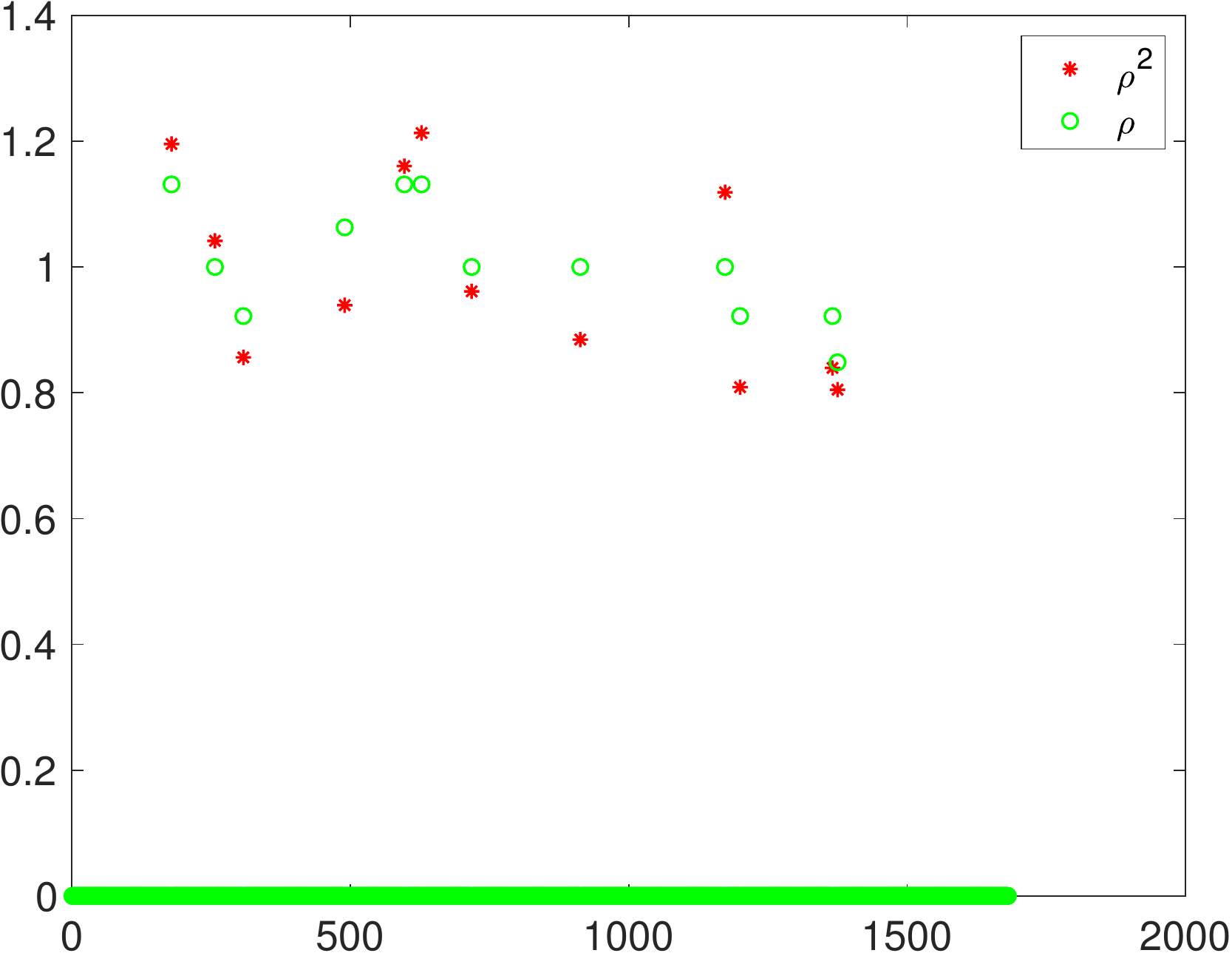}\\
\end{tabular}
\end{center}
\caption{High level of noise; SNR~$=1$. From left to right: $\ell_1$-norm minimization without the noise collector; 
$\ell_1$-norm minimization with a noise collector with $\Sigma=10^4$ columns, and $\tau=1$ in \eqref{eq:algo}; 
$\ell_1$-norm minimization with a noise collector, and the correct $\tau=2$ in \eqref{eq:algo}; $\ell_2$-norm  solution restricted to the support.
In the top row we show the images, and in the bottom row the solution vector with red stars and the true solution vector 
with green circles.}
\label{fig1}
\end{figure*}

In Fig. \ref{fig1}, we display the imaging results, with and without a Noise Collector, when the data is corrupted by additive noise. The SNR~$=1$, so the $\ell_2$-norms of the signals and the noise are equal. 
In the left plot, we show the recovered image using $\ell_1$-norm minimization without a Noise Collector. There is a lot of grass in this image, with many non-zero values outside the true support. 
When a Noise Collector is used, the level of the grass is reduced and the image improves; see the second from the left plot. Still, there are  several false discoveries  because we use $\tau=1$ in [\ref{eq:algo}]. 

In the third column from the left of Fig. \ref{fig1} we show the image obtained with a weight $\tau=0.8\sqrt{\ln 625}=2$ 
in [\ref{eq:algo}]. 
With this weight, there are no false discoveries and the recovered support is exact. This simplifies the imaging problem dramatically, as we can now
restrict the inverse problem to the true support just obtained, and then solve an overdetermined linear system using a classical $\ell_2$ approach. The results are 
shown in the right column of Fig. \ref{fig1}. Note that this second step largely compensates for 
the signal that was lost in the first step due to the high level of noise.

\begin{figure}[htbp]
\begin{center}
\includegraphics[scale=0.41]{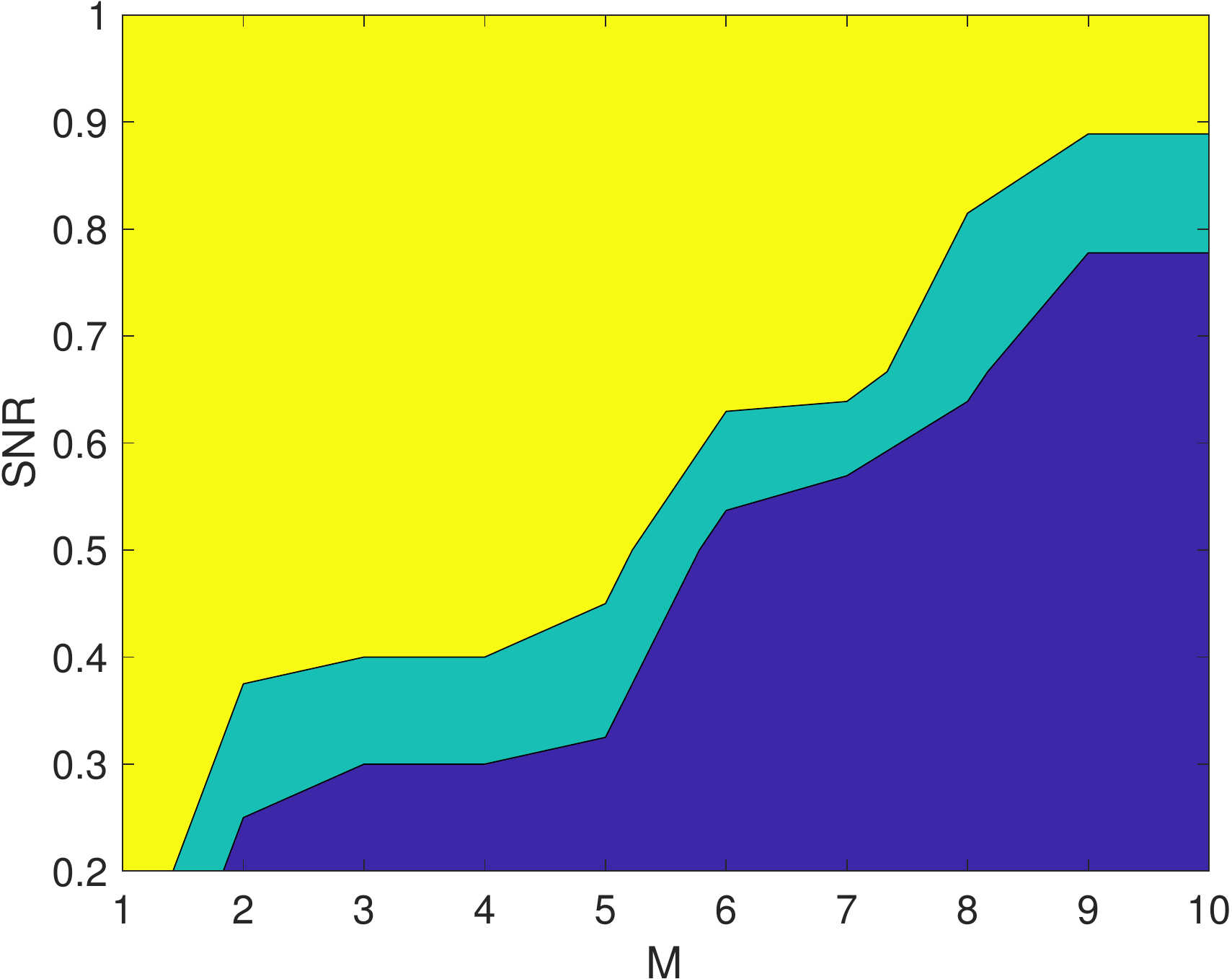} 
\end{center}
\caption{Algorithm performance for exact support recovery. Success corresponds to the value $1$ (yellow) and failure to $0$ (blue). The small phase transition zone (green) contains  intermediate values. Ordinate and abscissa are the sparsity $M$ and SNR.} 
\label{fig2}
\end{figure}

In Figure \ref{fig2} we illustrate the performance of the {\em Noise Collector} for different sparsity levels $M$ and SNR values. Success in recovering the true support of the unknown corresponds to the value $1$ (yellow) and failure to $0$ (blue). The small phase transition zone (green) contains  intermediate values. These results are obtained by averaging over 5 realizations of noise. 

Remark 1: We considered passive array imaging for ease of presentation. Same results hold for active array imaging with or without multiple scattering;
see \cite{CMP14} for the detailed analytical setup. 

Remark 2:
We have considered a microwave imaging regime. Similar results can be obtained in other regimes.

\vspace{-0.2cm}
\section*{Proofs}\label{sec:proofs}
\vspace{-0.15cm}

Proof of Lemma 1: Denote the event 
\[
\Omega_t=\left \{ \max_{i, j} |\langle \vect a_i, \vect c_j \rangle | \geq t/\sqrt{N}  \right\}.
\] 
By independence,
$\mathbb{P}\left(   |\langle \vect a_i,\vect c_j \rangle | \geq t/\sqrt{N} \right) \leq 2 \exp(- t^2/2)$ 
for any $i$ and $j$. Thus,
$\mathbb{P}\left( \Omega_t  \right) \leq 2 N \Sigma \exp(- t^2/2)$.
Choosing $t = c_0 \sqrt{\ln N}$ for sufficiently large $c_0$, we get 
\[
\mathbb{P}\left( \Omega_t   \right)  \leq 
C N^{\beta+1} N^{-c^2_0/2}\leq N^{-\kappa},
\]
where $c_0^2 > 2 (\beta +\kappa +1)$ and $N \geq N_0$. Hence, ~[\ref{deco_2}] holds with large probability $1-N^{-\kappa}$.

Next, we consider the chances that~[\ref{deco_1}] does not hold.  Suppose there is a direction $\vect b \in \mathbb{S}^{N-1}$ such that
\begin{equation}\label{anti_deco_1}
  |\langle \vect b,\vect c_j \rangle | \leq  \alpha/\sqrt{N}
\end{equation}
holds for all $j$.  Let 
$V_{k}(\vect c_{i_1}, \dots, \vect c_{i_k} )$
be the $k$-dimensional volume of a parallelogram spanned by  $\vect c_{i_1}$, $\dots$, $\vect c_{i_k}$. Note that $V_{k}$ is
equal to $V_{k-1}$ times its height. Then, if~\eqref{anti_deco_1} holds,
\begin{equation}\label{volumes}
\frac{V_{N}(\vect c_{i_1} , \dots, \vect c_{i_{N-1}},  \vect c_{i_N} )}{V_{N-1}(\vect c_{i_1},\dots, \vect c_{i_{N-1}} )} \leq \frac{2\alpha}{ \sqrt{N}}  
\end{equation}
for any choice of $N$ columns $\vect c_{i_j}$ from the noise collector $\cC$. If we fix  the indices 
${i_1}$, $\dots$, ${i_N}$ then,  due to rotational invariance, the probability of the event~\eqref{volumes}
equals the probability of event
$| \langle \vect c_1, \vect e_1 \rangle | \leq   2\alpha/\sqrt{N}$.  
Using 
\[
\mathbb{P} \left( | \langle \vect c_1, \vect e_1 \rangle | \leq   \frac{2\alpha}{ \sqrt{N}} \right) = \sqrt{\frac{N}{2\pi}}\int_{-2\alpha/\sqrt{N}}^{2\alpha/\sqrt{N}} \!\!\!\! e^{-x^2N/2} dx
\leq \frac{4 \alpha}{\sqrt{2 \pi}},
\]
and that we can find $N^{\beta -1}$ sets of distinct  indices  ${i_1}$, $\dots$, ${i_N}$, we conclude that
\[
\mathbb{P} \left(  \exists \vect b \in \mathbb{S}^{N-1} \hbox{ such that~\eqref{anti_deco_1} holds}    \right) \leq \left( 4 \alpha/\sqrt{2 \pi} \right)^{N^{\beta -1}}.
\]
Choosing $\alpha$ sufficiently small, i.e. $\alpha<4\sqrt{2 \pi}/4\approx 0.63$, and $N$ sufficiently large, we obtain the result.$\Box$

 Proof of Theorem~1:  In order to check~\eqref{no_signal_2}, we assume that both $\vect c_i$ and $-\vect c_i$ are in  $\Cc$, because it is more geometrically intuitive 
to work with the convex hull
\begin{equation}\label{hull}
H =\left\{ x \in \mathbb{R}^{N} \left| x=   \sum_{i=1}^{\Sigma} \xi_i \vect c_i ,~\xi_i \geq 0,~\sum_{i=1}^{\Sigma} \xi_i \leq  1  \right. \right\}.
\end{equation}
It implies we  may also assume $\vect \eta$ in~\eqref{rho_t} has non-negative coefficients,
 and $ \| \vect \eta \|_{l_1} = \min_{\lambda >0} \{ \vect e \in \lambda H \}$.
Thus,
 $\| \vect \eta \|_{l_1}$ is a norm of $\vect e$ with respect to $\Cc$, and  we can set  $ \| \vect e  \|_{\Cc} := \| \vect \eta \|_{l_1}$. This norm is called atomic in~\cite{recht10}.
 Suppose $\Lambda$ is the support of  $\vect \eta$. Its typical size $|\Lambda| = N$. 
Then, the simplex 
\[
\left\{ \vect x \in \mathbb{R}^{N} \left|  \vect x=    \sum_{i \in \Lambda} \alpha_i \vect c_i, \sum_{i \in \Lambda} \alpha_i =  1, \alpha_i \geq 0  \right. \right\}
\]
has the unique normal vector $\vect n$, which is collinear to our dual certificate $\vect z$, because
\begin{equation}\label{normal_}
\langle \vect z, \vect c_{i} \rangle  =\langle \vect z,  \vect e \rangle/\| e \|_{\Cc} =1,  \forall i  \in \Lambda, 
\langle \vect z, \vect c_{j} \rangle  < 1,  \forall j  \not\in \Lambda.
\end{equation}
 The estimate~\eqref{deco_1} implies that the convex hull $H$ contains an $l_2$-ball of radius $\alpha/\sqrt{N}$.  Therefore,
  $ \| \vect z\|_{l_2} \leq  \sqrt{N}/\alpha$  with large probability.

By construction, the distribution of $\Phi_\Cc(\vect e)$ is rotationally invariant with respect to the probability measure induced by all $\vect c_i$ and $\vect e$. Thus  
$\vect   n= \vect z/\| \vect z\|_{l_2}$
is also uniformly distributed on $\vect S^{N-1}$, and
\begin{equation}\label{expo}
\mathbb{P} \left(  \left| \langle \vect   a_j,  \vect n \rangle   \right| \geq t/\sqrt{N}  \right) \leq 2 \exp \left( -  t^2/2 \right),
\end{equation}
 for all $i=j,\dots,K$, see e.g.~\cite{vershynin}. Therefore, we can bound the probability that \eqref{no_signal_2} does not hold:
 \[
 \mathbb{P} \left( \max_{j \leq K}  |\langle  \vect a_{j}, \vect z \rangle | \geq  \tau \right) \leq K   \mathbb{P} \left( |\langle \vect a_{1}, \vect n \rangle | \geq  
 \alpha \tau/\sqrt{N} \right) 
 \]
 \[
\leq 2 K \exp \left( -  \alpha^2\tau^2/2 \right)  \leq  N^{\beta - \alpha^2 c^2_0/2} \sim N^{-\kappa},
 \]
 for large $N$ and appropriately chosen $c_0=\sqrt{2(\kappa + \beta)}/\alpha$. 
 Hence,  \eqref{no_signal_2} holds with large probability $1-N^{-\kappa}$. $\Box$

 Proof of Theorem~2: If the columns of $\Ac$ are orthogonal, our previous arguments could be modified to verify  Theorem 2. Indeed, suppose $V$ is the  span of the column vectors $\vect a_j$, with $j$ in the support of $\vect \rho$. Say,  $V$ is spanned by $\vect a_1$, $\dots$, $\vect a_{M}$. Let $W=V^{\perp}$ be the orthogonal complement to $V$. Then, the orthogonal projection of the signal $\vect \rho^{w} =0$. 
 By the concentration of measure see e.g.~\cite{vershynin},  the projection of the noise $\vect e^{w}$ is 
 uniformly distributed on  the unit sphere $\mathbb{S}^{N-1-M}$ with large probability. Applying the previous arguments to
  $\vect z^{w}$, the projection of $\vect z$ on $W$,  we conclude that
  the projection $\vect \rho^{w}_\tau=0$.  Therefore, $\mbox{supp}(\vect \rho_\tau) \subseteq \mbox{supp}(\vect \rho)$ with large probability.

For general $\Ac$ consider the orthogonal decomposition  $\vect a_i =\vect a_i^{v} + \vect a_i^{w}$ for all $i \geq M+1$.
As before, we can choose $\tau= c_0 \sqrt{ \ln N}$ so that
$ | \langle \vect a_i^{w}, \vect z \rangle | < \tau/2$ with large probability. It remains to demonstrate that
$| \langle   \vect a_i^{v}, \vect z \rangle | \leq \tau/2$. Fix any $i \geq M+1$.  
Suppose $\vect a_i^{v} = \sum_{k=1}^{M} \alpha_k \vect a_k$, and
$|\alpha_j| =\max_{k \leq M} |\alpha_k| = \| \vect \alpha \|_{l_\infty}$.
Thus,
 \[
\frac{1}{3M} \geq |\langle \vect a_j, \vect a_i^{v} \rangle | \geq  |\langle \vect a_j,  \sum_{k=1}^{M} \alpha_k \vect a_k \rangle | \geq   \| \vect \alpha \|_{l_\infty} \left( 1 -\frac{M-1}{3 M} \right).
\]
Then, $\| \vect \alpha \|_{l_\infty} \leq 1/2M$, so $\| \vect \alpha \|_{l_1} \leq M \| \vect \alpha \|_{l_\infty} \leq 1/2$. Hence,
\[
| \langle  \vect a_i^{v}, \vect z \rangle | \leq  \sum_{k=1}^{M} |\alpha_k|~|\langle \vect a_k, \vect z \rangle| \leq  \| \vect \alpha \|_{l_1}  \tau \leq \tau/2.  \hspace{1.5cm} \Box
\]

 Proof of Theorem~3:  It suffices to prove the result for 1-sparse $\vect \rho$, say,   $\vect \rho =(1, 0, \dots, 0)$.
We will demonstrate that the solution to the minimization problem 
\begin{align} \label{lastStep}
\left( \vect \eta_{\tau}, \rho_1 \right) = \arg\min_{\vect \rho, \vect \eta} 
\left( \| \vect \eta \|_{\ell_1} +  \tau |\rho| \right), \\ \nonumber
\hbox{ subject to }  \cC \vect \eta = \vect e + \vect a_1 (1- \rho_1),
\end{align}
with $\tau=c_0\sqrt{\ln N}$, satisfies $\rho_1 >1/2$ if $\delta < c_2/\sqrt{\ln N}$, $c_2=\alpha/5 c_0$. This implies $\mbox{supp}(\bfrho_\tau) = \mbox{supp}(\bfrho)$.

Suppose $\vect \eta_{\vect e}$,  $\vect \eta_{\vect a_1}$ and $\vect \eta_{t}$ are the optimal solutions of
 \begin{equation}\label{minimax_c}
\vect \eta_{\vect b} = \arg\min \left(   \| \vect \eta \|_{\ell_1} \right), \hbox{ subject to } \cC \vect \eta = \vect b,
 \end{equation}
with right-hand sides $ \vect b= \vect e$,  $\vect b= \vect a_1$, and    $\vect b=\vect e + t \vect a_1$, respectively.
Since $ \cC \left( \vect \eta_{t} - \vect \eta_{\vect e} \right) = t \vect a_1$, we have
\[
t \| \vect \eta_{\vect a_1} \|_{\ell_1} \leq \| \vect \eta_t \|_{\ell_1} + \| \vect \eta_{\vect e} \|_{\ell_1}\, 
\]
and, therefore,
\begin{equation}\label{t_bound}
\| \vect \eta_t \|_{\ell_1}  \geq t \| \vect \eta_{\vect a_1} \|_{\ell_1}-  \| \vect \eta_{\vect e} \|_{\ell_1}\, .
\end{equation}

From~\eqref{deco_1} and~\eqref{l1_bound}, we have
 \[
 \| \vect \eta_{\vect e} \|_{\ell_1} \leq \delta \frac{\sqrt{N}}{\alpha} \hbox{ and }  \| \vect \eta_{\vect a_1} \|_{\ell_1} \geq  \frac{\sqrt{N}}{c_0 \sqrt{\ln N}},
 \]
 respectively. 
Suppose $\delta \leq c_2/\sqrt{\ln N}$, $c_2=\alpha/5 c_0$. Then, for any 
$t  \geq 1/2$, and for $N$ large enough,
 \[
 t \| \vect \eta_{\vect a_1} \|_{\ell_1} - \| \vect \eta_{\vect e} \|_{\ell_1} >  \| \vect \eta_{\vect e} \|_{\ell_1} + \tau t.
 \]
 Using~\eqref{t_bound} with $t=1-\rho_1$,  we conclude that
 \[
 \| \vect \eta_{1-\rho_1} \|_{\ell_1}  +  \tau \rho_1 >  \| \vect \eta_{\vect e} \|_{\ell_1} +\tau
 \]
 for all  $\rho_1 \leq 1/2$. It implies \eqref{lastStep}. 
$\Box$

{\bf Acknowledgements} 
The work of M. Moscoso was partially supported by Spanish grant MICINN FIS2016-77892-R. The work of A.Novikov was partially supported by NSF grants  DMS-1515187, DMS-1813943.
The work of G. Papanicolaou was partially supported by AFOSR FA9550-18-1-0519. The work of C. Tsogka was partially supported by AFOSR FA9550-17-1-0238 and FA9550-18-1-0519. We thank Marguerite Novikov for drawing Figure~\ref{phi_map}.


\end{document}